# Improved properties of epitaxial $YNi_xMn_{1-x}O_3$ films by annealing under high magnetic fields

Yanwei Ma [a)], Aixia Xu, Xiaohang Li, Xianping Zhang

*Institute of Electrical Engineering, Chinese Academy of Sciences, P. O. Box 2703, Beijing 100080, China*

M. Guilloux-Viry, O. Peña

*Unité de Sciences Chimiques de Rennes, UMR 6226, CNRS - Université de Rennes 1 - 35042 Rennes cedex, France*

S. Awaji, K. Watanabe

*High Field Laboratory for Superconducting Materials, Institute for Materials Research, Tohoku University, Sendai 980-8577, Japan*

**Abstract**

The effect of annealing under a magnetic field on the microstructure and properties of $YNi_xMn_{1-x}O_3$ (x=0.33 and 0.5) films has been investigated. It is found that the ferromagnetic transition temperature is significantly enhanced after post annealing in the presence of an 8 T magnetic field. Characterization study shows that the microstructure is affected, obtaining larger grains of uniform size when films are annealed under a magnetic field. The improvement in the ordering temperature of all films is interpreted in terms of the grain growth caused by the magnetic-field driving force for boundary motion where the exchange coupling is high.

---
[a)] Electronic mail: ywma@mail.iee.ac.cn

YMnO$_3$ is a ferroelectric, antiferromagnetic compound of hexagonal structure. The substitution of Ni$^{2+}$ for Mn$^{3+}$ leads to a phase transition from hexagonal towards an orthorhombic perovskite phase for Ni amounts larger than 20 at.%; At the other end, the solid solution is limited to a maximum of 50 at.% Ni, which means that Ni adopts a stable divalent state [1]. The magnetic properties of the YNi$_x$Mn$_{1-x}$O$_3$ solid solution change with the Ni content, evolving from an antiferromagnetic behavior at low nickel concentration to ferromagnetism at high nickel content, the threshold value being the critical concentration x(Ni) = 1/3 [2]. It becomes then interesting to investigate such system in thin-film form since epitaxial growth may improve the conducting properties and eventually, disable the phase segregation which may appear at the frontier of antiferromagnetism and ferromagnetism. Recently, we have grown epitaxial Y(Ni,Mn)O$_3$ (YNMO) thin films on SrTiO$_3$ substrates by pulsed laser-ablated deposition techniques and investigated the magnetic and microstructural properties [3].

On the other hand, high magnetic fields are known to modify the microstructure of materials during their fabrication process. In most cases, a weak paramagnetic magnetization is usually coupled to a strong field to orientate anisotropic grains [4-6]. It has been reported that the microstructure and magnetic properties of manganite films are sensitive to preparation and annealing conditions [7-8]. Stronger effects or unknown effects are expected if applying a high magnetic field during YNMO post deposition annealing, a process which we will refer hereafter as "magnetic annealing". In the present work, we have investigated the effect of such magnetic annealing on thin films of YNi$_x$Mn$_{1-x}$O$_3$ grown on SrTiO$_3$ substrates. For this, we have chosen two characteristic compositions, x=0.33 and 0.5, because they better characterize the magnetic and electrical behavior of this series, as exposed above. In the course of our investigations we have found that magnetic field annealing effectively promotes the ordering temperature of YNMO films.

Thin films of YNMO were grown using pulsed laser deposition. Deposition was performed from targets with stoichiometric composition of YNi$_x$Mn$_{1-x}$O$_3$ (x=0.33, 0.5). Films were synthesized on SrTiO$_3$ (100) substrates (cubic, a=3.905 Å). A detailed description of the deposition system is mentioned elsewhere [9]. In brief, a 248 nm KrF pulsed laser with 2 Hz repetition rate and 2 J/cm$^2$ energy density was used. Deposition was performed at 740 °C under an oxygen pressure of 0.25-0.6 mbar. Following the deposition, the films were cooled down to room temperature at a rate of about 35 °C/min in 200 Torr of oxygen. All films had a thickness around 200 nm.

Then, the samples were annealed with a flowing oxygen gas in an electrical furnace, which was installed in the room temperature bore of a cryogenfree superconducting magnet [10]. A magnetic field of 8 T was applied during annealing. The annealing temperature was 850°C; the annealing time was 2-10 h. In order to evaluate the effect of the magnetic field, the samples were annealed under the same conditions but without any magnetic field. The direction of the magnetic field was perpendicular to the film plane.

The structure of the films was examined by X-ray diffraction (XRD). The microstructure was observed using a field effect scanning electron microscopy (SEM). Field-cooled (FC) and zero-field-cooled (ZFC) magnetization was measured at various applied fields in a superconducting quantum interference device magnetometer. The applied field was in the film plane. It should be noted that for the same composition of films, we have used exactly the same specimen to make the comparison, for instance, the sample annealed under a magnetic field is the same as the as-grown one; then their thickness should be equivalent. For this reason, magnetization is just given in arbitrary units (emu), without normalization to a sample's parameter (mass or volume).

XRD patterns of YNMO films (x=0.5 and 0.33) from the as-grown to the different anneals are shown in Fig.1. For the as-deposited films, only 00$l$ diffraction peaks of the phase are present evidencing an oriented film with the c-axis perpendicular to the surface of the substrate. XRD patterns of non-field and magnetic field annealed films were similar to that of as-deposited films. However, there is a significant increase in the intensity of diffraction peaks as the samples were annealed, especially in the case of magnetic field annealing. Crystalline quality of the films was analyzed using the measured full width at half maximum (FWHM) of the rocking curves. X-ray rocking curves of the (004) peak show that the crystalline mosaic spread decreases with different annealing for both x = 0.33 and 0.5 films. As for the x=0.33 films, values of FWHM are 1.02, 0.32 and 0.25° for the as-grown, non-field and magnetic field anneal states, respectively. This compares with a change of 1.16, 1.1 to 0.8° for the x=0.5 films from the as-grown, non-field to magnetic anneal states. These results suggest that the crystalline quality is significantly improved by application of the external field in comparison with the as-grown ones. It is also clear that magnetic field annealing is much more effective in enhancing the crystallinity of the films compared to non-field heat treatments.

SEM observations further demonstrate that magnetic field annealing has an effect on the microstructural evolution. Figure 2 shows the morphology of as-grown, non-field and field-annealed YNMO films for x= 0.33 and 0.5. After the thermal treatment, the grain shape seems similar to that of as-grown films, but the grain size is largely increased and the boundaries between grains become blurred. Clearly, upon the magnetic annealing, the grains in the YNMO films were significantly enlarged and the grain boundary density was consequently reduced. In the case of x=0.5, the as-grown film consists of in-plane oriented longitudinal islands with an average grain size of 60 nm (Fig.2d). However, the films subjected to magnetic field anneal show significant increase in the grain size (up o145 nm), more than two times larger than that of the as-grown film while slight increment of grain size was observed in the non-filed annealed films (Fig.2e). Similarly, for x=0.33, the films annealed in the field also show remarkable enhancement in the grain size (108 nm, Fig.2c) compared to the as-deposited film (spherical grains with an average grain size of 30 nm, Fig.2a). The average grain sizes for all the films are summarized in Table I. Clearly, magnetic annealing of the YNMO film results in not only a grain growth which reduces the density of grain boundaries, but also an improvement in the film crystallinity, as supported by the XRD measurements, thus increasing the exchange coupling. Therefore, enhanced magnetic properties are expected after magnetic annealing.

The ZFC/FC magnetization of as-grown and magnetic annealed YNMO films is shown in Fig. 3. As-grown films of x=0.5, typically, have a ferromagnetic transition temperature Tc ~ 85 K; magnetic annealing raises Tc to about 93 K. In x=0.33 films, the as-grown Tc is ~60 K. Magnetic annealing causes a dramatic improvement in the properties of YNMO films. As shown in Fig.3a, in annealed films we observe Tc~80 K, that is, an increase of Tc by ~20 K is achieved, bringing the transition in this composition approximately equal to the transition temperature of x=0.5 as-grown films. At the same time, the spin canting-like transition $T_{max}$ (defined at the maximum value of the ZFC magnetization) increases from about 42 to 48 K. Along with the enhanced Tc, DC magnetization measurements at low fields (as exemplified in Fig.3) indicate a remarkable increase in the field-cooling magnetic moment in all the samples after magnetic annealing. This increase of the total magnetic moments strongly suggests an increase in the FM moments due to magnetic annealing. These clearly show that the magnetic annealing is helpful to improve the magnetic properties of YNMO films.

The effect of magnetic field annealing is further demonstrated by the field dependence of magnetization measurements for x = 0.33 and 0.5 films at 5 K as shown in Fig.4. As for the x=0.5 film subjected to magnetic annealing, the magnetization not only is much larger, but also saturates more easily than that of as-grown samples (Fig.4b), indicating a more typical FM character. Similarly, magnetically annealed films of x=0.33 show also a considerable increase of the magnetization. However, magnetic annealing hardly changes the coercive field, keeping a value of the order of 1500 Oe for both x=0.5 and 0.33 films.

It should be noted that, as reported in our previous work [11], annealing without an external magnetic field just shows a small enhancement by 10 K in Tc for the films with x=0.33 while little effect on the Tc is observed for the x=0.5 films. Such a slight increase of the transition temperatures Tc under non-field annealing is not surprising since the post deposition annealing can lead to an increase of the oxygen content of the films, optimizing the ratio $Mn^{3+}/Mn^{4+}$ [7-8, 12]. As we see, the enhancement of Tc resulting from non-field annealing is quite small compared to the magnetic field annealing process, as presented in Table 1. However, Tc of the x=0.5 samples subjected to magnetic annealing has increased by 8 K but almost no change in Tc was observed using non-field annealing. It thus demonstrates that a remarkable improvement of magnetic properties of the field-annealed films can be clearly attributable to the effects of the external magnetic field.

From the above results, not only the crystallinity becomes improved, but also the grain size is increased when the magnetic field is applied to annealing process for YNMO films. It is also recognized that the larger the grain, the lower the density of grain boundaries. Thus the exchange coupling between grains is enhanced and hence the Tc improvement. As reported earlier, Tc decreases when the grain size decreases, eventually giving rise to superparamagnetic particles without a distinct Tc when the particle size is very small [13]. The fact also corroborated by the results of Nam et al.[14] They found the better crystalline quality and good grain coupling can lead to the better physical properties. Clearly, our observation of Tc enhancement by magnetic annealing is in good agreement with this viewpoint. Therefore, the mechanism operating to achieve such improvement of magnetic properties is closely related to better crystallinity and reduction of grain boundaries caused by magnetic annealing.

The question now is how does the magnetic field influence the grain growth and grain boundary by magnetic sintering? It has been found [15-16] that an applied magnetic field during annealing can play a significant role even when the material is in its paramagnetic state above the Curie point, as pointed out by Tsurekawa et al. in iron samples [15]. In our case, YNMO is in its paramagnetic state at 850°C. Applying a magnetic field forces the magnetic moments to align in the direction of the external field, in a much more efficient way than a random disorientation due to thermal motion. This magnetic ordering together with magneto-crystalline anisotropy provides a driving force for grain boundary migration that greatly contributes to the grain growth [17]. As a consequence, grains become larger, and crystallinity is enhanced, making the exchange-coupled adjacent grains crystallographically coherent, thus improving the magnetic properties of the YNMO films.

In summary, it is found that annealing under a magnetic field leads to a significant improvement in several magnetic properties of both x=0.33 and 0.5 films of the $YNi_xMn_{1-x}O_3$ system. In particular, we observe that the ferromagnetic transition occurs at much higher temperatures compared to the corresponding films subjected to the ordinary sintering. The enhanced properties may be explained by improved crystallinity and reduced density of grain boundaries, which is caused by magnetic field annealing.

The authors would like to thank Heying Wang, Li Han, Liye Xiao, P. Badica, T. Nojima for their help. C. Moure, of the CSIC-Spain, is greatly acknowledged for providing the YNMO targets used for laser-ablation deposition. This work is partially supported by the National Science Foundation of China (NSFC) under Grant No.50472063.

**Table**

TABLE I.  Transition temperatures (Tc) and average grain size of x=0.33 and 0.5 films of as-grown, non-field annealing and magnetic field annealing

| Films | As-grown | | Non-field annealing | | Magnetic field annealing | |
|---|---|---|---|---|---|---|
| | Tc | Grain size | Tc | Grain size | Tc | Grain size |
| x=0.33 | 60 K | 30 nm | 70 K | 75 nm | 80 K | 108 nm |
| x=0.5 | 85 K | 60 nm | 85.5 K | 80 nm | 93 K | 145 nm |

**Captions**

Figure 1  XRD patterns of $YNi_xMn_{1-x}O_3$ films subjected to different annealing conditions. For x=0.33: (a) as-gown, (b) non-field annealing, (c) magnetic field annealing; For x=0.5: (d) as-gown, (e) non-field annealing, (f) magnetic field annealing. Note: two small peaks at 2θ=38.286° and 44.423° were contributed by the sample holder.

Figure 2  Images of typical areas of $YNi_xMn_{1-x}O_3$ films. For x=0.33: (a) as-gown, (b) non-field annealing, (c) magnetic field annealing; For x=0.5: (d) as-gown, (e) non-field annealing, (f) magnetic field annealing.

Figure 3  Thermal variation of the ZFC/FC magnetization for the following $YNi_xMn_{1-x}O_3$ films: (a) x = 0.33 and (b) x = 0.5. The inset shows the enlarged view near the transition temperature. Note that the as-grown sample was measured and then subsequently magnetically annealed to give the 'magnetically annealed' sample.

Figure 4  Hysteresis loops, up to 3 T, measured at 5 K on $YNi_xMn_{1-x}O_3$ films of x = 0.33 and x = 0.5. Note that the as-grown sample was measured and then subsequently magnetically annealed to give the 'magnetically annealed' sample.

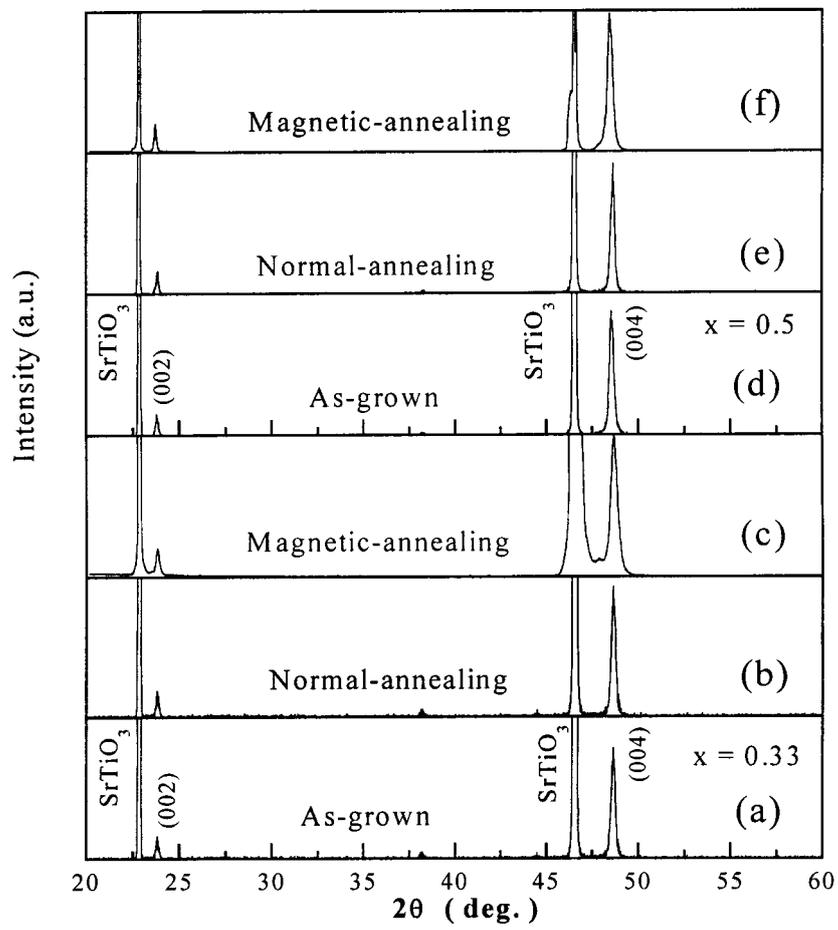

Fig.1

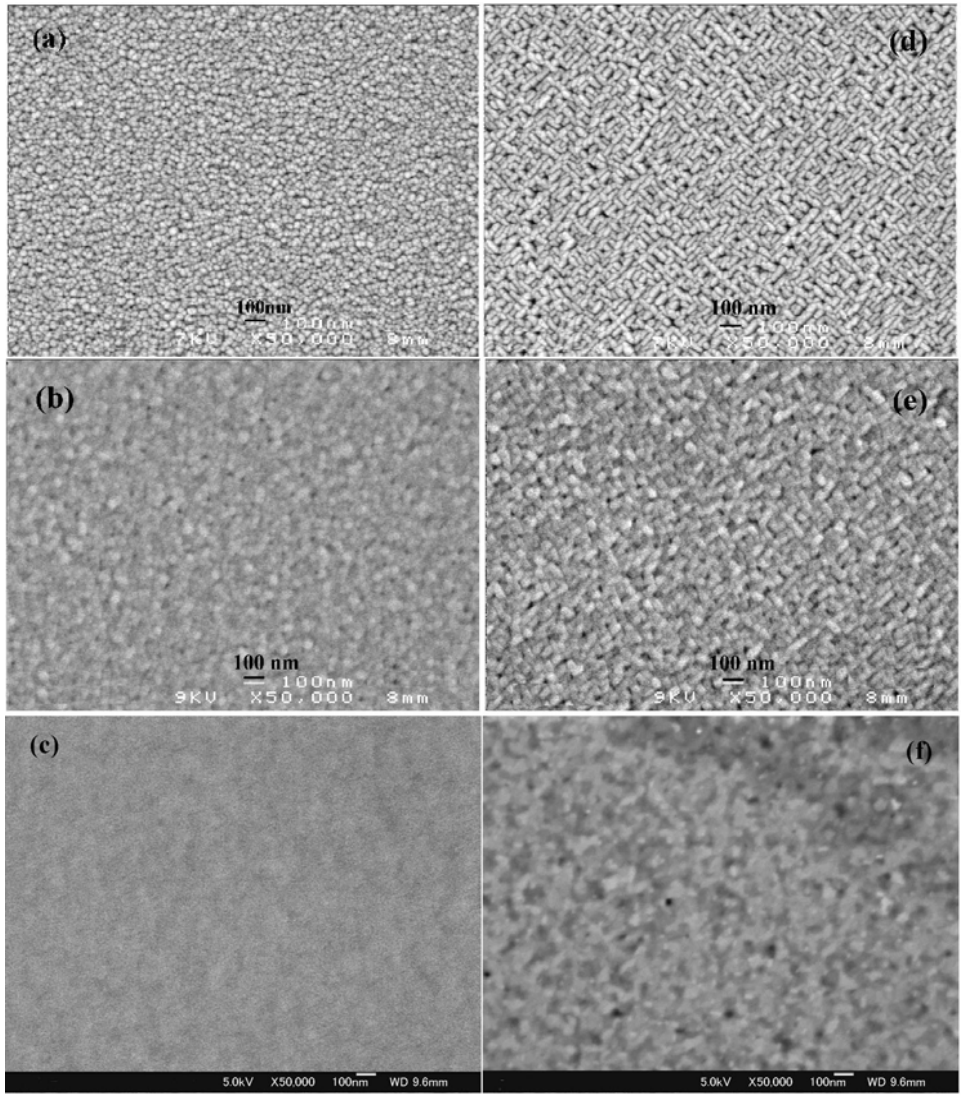

Fig.2

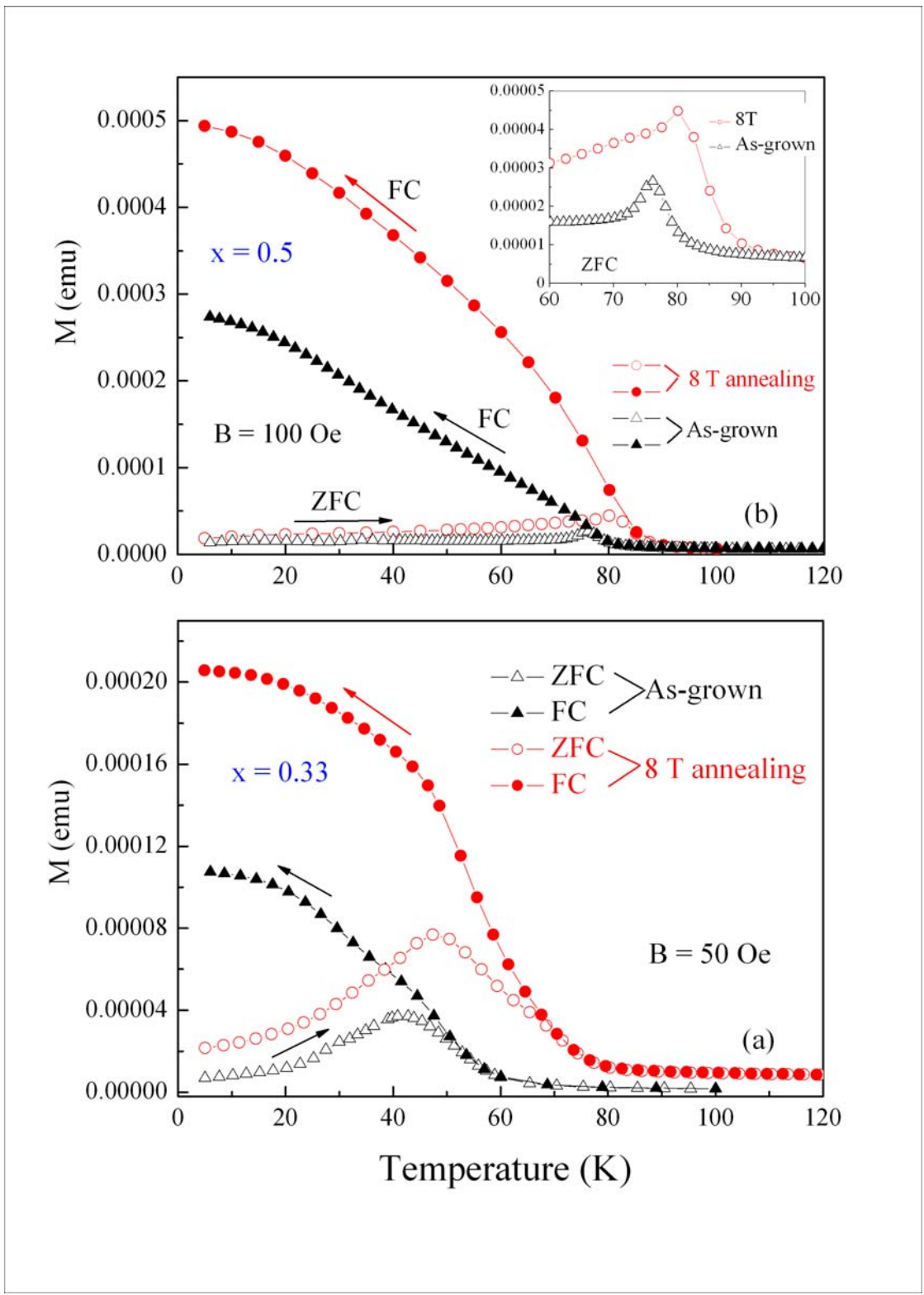

Fig.3

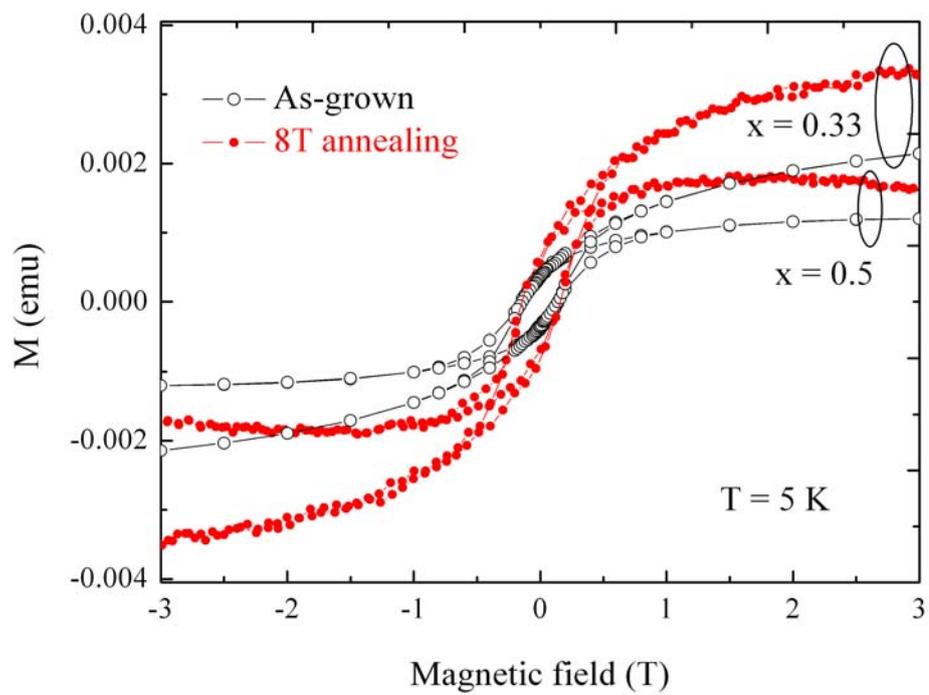

Fig.4